\documentclass[sigconf, 9pt, nonacm]{acmart}
\usepackage{enumitem}

\usepackage{hyperref}
\usepackage{subcaption}
\usepackage{breqn}
\usepackage[dvipsnames]{xcolor}
\usepackage{caption}   %
\usepackage[table]{xcolor} %
\usepackage{colortbl}      %
\usepackage[acronym]{glossaries} %
\makeglossaries

\newlength{\tableonecolwidth}
\newlength{\tabletwocolwidth}
\newcommand{\sysname}{WiNeRF}

\newcommand{\constraints}{measurement constraints }
\newcommand{\wavemodel}{Cone Sampling Wave Model }
\newcommand{\worldmodel}{Implicit Scene Representation }
\newcommand{\rendering}{CSI Reconstruction Module }
\newcommand{\rev}[1]{\textcolor{black}{#1}}

\newacronym{los}{LoS}{line-of-sight}
\newacronym{nlos}{NLoS}{non-line-of-sight}
\newacronym{csi}{CSI}{channel state information}
\newacronym{aoa}{AoA}{angle-of-arrival}
\newacronym{isac}{JSAC}{Joint Sensing and Communication}
\newacronym{nerf}{NeRF}{Neural Radiance Field}
\newacronym{rssi}{RSSI}{received signal strength}
\newacronym{rx}{Rx}{receiver}
\newacronym{cfo}{CFO}{carrier frequency offset}
\newacronym{sfo}{SFO}{sampling frequency offset}
\newacronym{ula}{ULA}{Uniform Linear Array}

\usepackage[subtle]{savetrees}
\hypersetup{pdfstartview=FitH,pdfpagelayout=SinglePage}

\newcommand{\saif}[1]{\textcolor{Black}{#1}}
\newcommand{\Raf}[1]{\textcolor{Black}{#1}}

\AtBeginDocument{%
  }

\setcopyright{acmlicensed}
\copyrightyear{2018}
\acmYear{2018}
\acmDOI{XXXXXXX.XXXXXXX}
\acmISBN{978-1-4503-XXXX-X/2018/06}

\begin{document}

\title{WiNeRF: \saif{Measurement Constrained} Radiance Fields for Actionable Wireless Channel Modeling}

\author{%
Saif Ur Rahman$^{1}$, 
Rafid Umayer Murshed$^{1}$, 
Anton Dmitriev$^{1}$, 
Cagri Tanriover$^{2}$, \\
Rahul C. Shah$^{2}$, 
Elahé Soltanaghai$^{1}$
}

\affiliation{%
  \institution{$^{1}$University of Illinois Urbana-Champaign, $^{2}$Intel}
  \country{}
}

\email{{saifu2,rum3,antond2,elahe}@illinois.edu, {cagri.tanriover,rahul.c.shah}@intel.com}

\renewcommand{\shortauthors}{Ur Rahman et al.}

\begin{abstract}

Wireless embedded systems increasingly rely on wireless channel information for decision making, yet practical platforms operate under severe constraints, including few antennas, narrow bandwidth, and sparse, noisy measurements. While neural field based approaches inspired by \gls{nerf} have recently been explored for continuous wireless channel modeling, existing approaches depend on dense measurements or external priors such as known geometry, visual context, or \gls{aoa} information, limiting their practicality in real-world deployments. We present \sysname{}, a neural field framework that learns a spatially continuous, complex-valued wireless channel representation directly from sparse \gls{csi} collected by commodity WiFi devices. \sysname{} embeds intrinsic system constraints, such as antenna geometry, limited spatial resolution, and phase uncertainty, as inductive biases through a 3D conical wave sampling model, a multi-resolution implicit scene representation, and a differentiable optimization framework for complex-valued channel learning. Across diverse indoor environments with \gls{nlos} regions, \sysname{} achieves a median prediction SNR of 5.3 dB, outperforming prior neural baselines by 4.9 dB on average (approximately 3× higher prediction SNR), and produces a task-agnostic channel representation that can be directly reused in standard signal-processing pipelines, including beamforming, \gls{aoa} estimation, and RSSI coverage mapping, without modifying existing hardware or wireless protocols.
\end{abstract}



\maketitle

\section{Introduction}

Modern embedded wireless systems increasingly seek to exploit wireless channel information to support environment-aware decision making under tight sensing and communication constraints. In these systems, communication signals can be \rev{leveraged} to sense and map the surrounding environment, enabling proactive decision-making that accounts for wireless conditions alongside physical factors \cite{intro_ISAC,intro_ISAC_2,intro_wifi_sensing}. For example, a mobile robot can map and avoid communication dead zones to plan connectivity-aware paths, while an embedded node can predict coverage gaps to maintain reliable links under mobility. Achieving this intelligence requires modeling the channel under severe measurement constraints, a task central to building high-fidelity Digital Twins and site-planning tools where scene-specific accuracy is the primary objective. In this context, \gls{csi} provides a rich representation of how wireless signals propagate and interact with surrounding objects \cite{intro_wifi_sensing}. Specifically, the phase variations of \gls{csi} across frequency subcarriers and antenna elements capture intrinsic multipath characteristics that define the environment's spatial structure \cite{kotaru2015spotfi}. As such, accurately capturing these multidimensional relationships is essential for understanding and predicting the static spatial behavior of the wireless channel in resource-constrained \gls{isac} systems.
\begin{figure}[tb]
\centering
\includegraphics[width=1.0\linewidth]{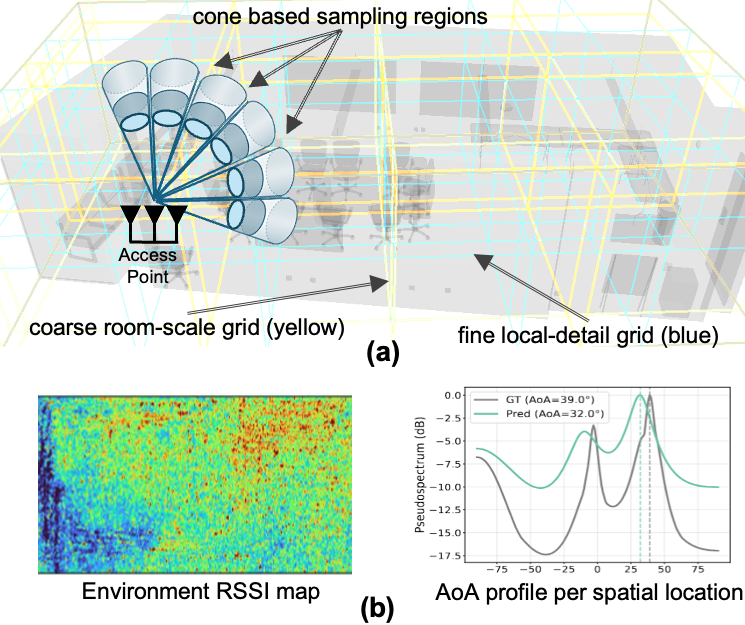}
\caption{\rev{(a) \sysname{} combines cone-based sampling regions with a multi-resolution implicit scene representation of coarse room-scale grids (yellow) and fine local-detail grids (blue) to learn a continuous wireless field from sparse \gls{csi} measurements. (b) The learned field supports downstream tasks such as RSSI coverage mapping and \gls{aoa} profiling.}}
\label{fig:concept_first}
\end{figure}

Prior works on wireless channel modeling have explored \gls{nerf}-based approaches to represent \saif{implicit volumetric representations} of wireless channels, capturing spatial signal spectra or coverage patterns \cite{zhao2023nerf2, chen2024rfcanvas, lu2024deep, jiang2025learnable}. However, all of these methods make strong assumptions such as dense training data \cite{zhao2023nerf2} or rely on priors such as known 3D geometry \cite{orekondy2023winert}, visual context \cite{chen2024rfcanvas}, or precise \gls{aoa} information \cite{lu2024deep} to guide learning and stabilize training. \Raf{This reliance makes them impractical for real-world deployments where such external data is often unavailable.} A key challenge is that wireless channel modeling is an \Raf{inherently difficult reconstruction task}, with more unknown variables (e.g. multipath characteristics) than available observations due to the small number of antennas and aggregate representation of multipath in the form of CSI matrices. \saif{Due to these reasons, prior works focus on task-specific objectives, rather than learning general-purpose channel representations that can be reused across downstream signal-processing tasks.}

To address these limitations, we ask a different question:
\textbf{\emph{Can we design a neural wireless channel model that uses measurement \saif{\constraints} as natural \Raf{guides} to produce actionable and \Raf{consistent} channel representations?}} \Raf{These intrinsic measurement characteristics, such as spatial resolution and antenna array geometry, fundamentally shape how the channel is observed at the wireless node. Rather than treating them as noise to be compensated for, we use them as inductive biases that regularize what the model is allowed to represent.}

\begin{figure}[t]
\centering
\includegraphics[width=1.0\linewidth]{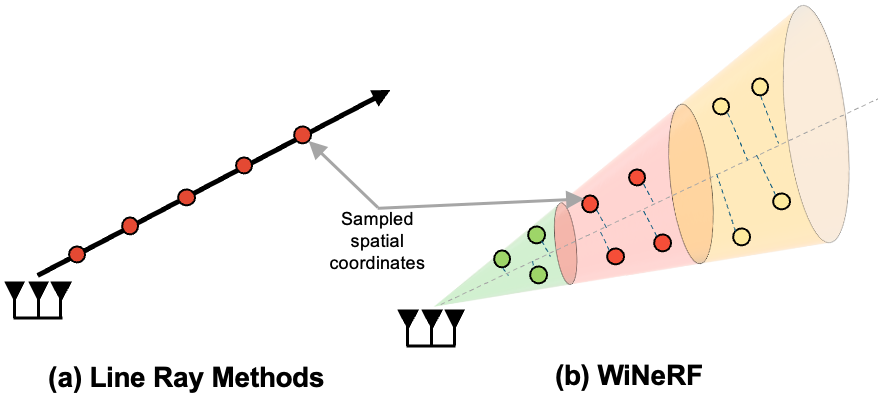}
\caption{\rev{(a) Prior NeRF-based wireless models sample along infinitesimally thin line rays, implicitly assuming arbitrarily fine angular resolution. In contrast, (b) \sysname{} replaces line rays with cone-based support regions that reflect finite antenna-array observability, enabling hardware-aware spatial aggregation.}}
\label{fig:concept}
\end{figure}

\rev{\saif{We present \sysname{}, a site-specific neural field framework for learning spatially continuous, complex-valued wireless channel representations subject to measurement constraints of commodity hardware}.} An overview of \sysname{} is presented in Figure.~\ref{fig:concept_first}. Prior NeRF-based wireless channel models often idealize propagation using infinitesimally thin rays or fixed-resolution voxel grids, implicitly assuming arbitrarily fine angular resolution at the receiver. In contrast, \sysname{} replaces line-ray sampling with conical primitives whose angular extent reflects the finite \gls{aoa} resolvability of practical antenna arrays, as illustrated in Figure.~\ref{fig:concept}. Importantly, these cones do not model directional radiation; rather, they encode the set of propagation directions that are indistinguishable given the hardware's angular observability, mitigating \gls{aoa} aliasing while preserving multipath-related structure.

\rev{As a result, \sysname{} learns a site-specific channel field that can be queried at unseen receiver locations within the same environment, even under lower measurement density than prior approaches.} \Raf{These estimates are immediately actionable for offline tasks such as beamforming, \gls{aoa} estimation, and spatial \gls{rssi} coverage mapping.} The framework consists of three interconnected measurement-driven modules: \emph{(i)~\wavemodel, (ii)~\worldmodel, and (iii)~\rendering}.

\vspace{.3em} \noindent  \textbf{\wavemodel:}  
Learning a continuous neural representation directly from sparse CSI is \Raf{difficult because many different signal paths can produce the same measurement at the receiver}. Instead of \Raf{relying on extra data like 3D maps or images}, we \Raf{guide} the representation using the \Raf{actual} measurement characteristics of the antenna arrays themselves. Specifically, we represent each propagation direction using a volumetric cone \Raf{that reflects} the array’s \Raf{natural} angular resolution, and approximate each cone with a multivariate Gaussian to enable stable, differentiable CSI rendering.
\saif{This replaces idealized line-ray sampling with a representation that explicitly accounts for the finite angular resolution of practical antenna arrays. The cone width encodes antenna-dependent angular ambiguities~\cite{matter2022ambiguities, goldsmith2005wireless}, such as left--right aliasing in linear arrays or elevation ambiguity in horizontal arrays. By aligning the sampling process with the receiver’s angular observability, the learned channel representation captures only multipath structure that is distinguishable under the available measurements.}

\vspace{.3em} \noindent \textbf{\worldmodel:} Unlike camera images where each pixel corresponds to a localized point in space \cite{szeliski2022computer}, a CSI measurement aggregates contributions from many multipath signals with different path lengths and angles; nearby objects can dominate certain measurements while distant reflectors may only appear at coarse spatial scales \cite{adib2013see, goldsmith2005wireless}. As a result, a continuous field model must capture meaningful world structure without over-smoothing or overfitting. This need is amplified because our cone formulation represents distributions of feasible arrival directions rather than single paths. 
\Raf{To address this, \sysname{} adopts a Multi-Scale Scene Representation using grids of various resolutions to represent the wireless environment. This compact and adaptive feature field is capable of encoding large room-scale boundaries (e.g., walls) and fine local reflectors (e.g., furniture) within the same framework. This multi-scale encoding preserves the spatial detail needed to separate complex signal paths while still generalizing from sparse \gls{csi}.}

\begin{table*}[t]
\centering
\caption{Evolution of Neural Field Methods for Wireless Applications}
\label{tab:evolution}
\resizebox{\textwidth}{!}{%
\begin{tabular}{|l|l|c|c|l|}
\hline
\textbf{Method} & \textbf{Wave Representation} & \textbf{Priors Need} & \textbf{Experimental Data} & \textbf{Channel Output} \\ \hline
\textbf{NeRF\textsuperscript{2}} & Voxelized Rays & Dense Training Data & \cellcolor[HTML]{FFFFFF}{\color[HTML]{009901} \checkmark} & Task-dependent (e.g. RSSI, spatial spectrum) \\ \hline
\textbf{WiNeRT} & Ray segments & Path parameters (AoA, ToF) &  \cellcolor[HTML]{FFFFFF}{\color[HTML]{FE0000} x}  & CIR \\ \hline
\textbf{RF-Canvas} & Tensorial Field & 3D Geometry & \cellcolor[HTML]{FFFFFF}{\color[HTML]{009901} \checkmark} & RSSI \\ \hline
\textbf{NeWRF} & Voxelized Ray & Multipath AoA &  \cellcolor[HTML]{FFFFFF}{\color[HTML]{FE0000} x}  & Complex CSI \\ \hline
\cellcolor[HTML]{FFFFFF}{\color[HTML]{009901} \textbf{WiNeRF}} & \cellcolor[HTML]{FFFFFF}{\color[HTML]{009901} \textbf{Hash-based Cones}} & \cellcolor[HTML]{FFFFFF}{\color[HTML]{009901} \textbf{--}} & \cellcolor[HTML]{FFFFFF}{\color[HTML]{009901} \textbf{\checkmark}} & \cellcolor[HTML]{FFFFFF}{\color[HTML]{009901} \textbf{Complex CSI}} \\ \hline
\end{tabular}%
}
\end{table*}

\vspace{.3em} \noindent \textbf{\rendering:}
This module combines the \wavemodel and \worldmodel to synthesize the \gls{csi} at \gls{rx} locations via differentiable rendering over cone-based primitives, producing complex-valued CSI across antennas and subcarriers.
A key challenge is that CSI phase from commodity hardware is heavily corrupted by noise and impairments such as phase wrapping, CFO, and SFO \cite{liu2021fire, tan2006adaptive}, making direct prediction of absolute phase unreliable. We therefore decouple hardware impairment compensation from learning by applying magnitude normalization, removing global CFO/SFO, and training with a relative phase loss that captures phase relationships across antennas and subcarriers. This preserves multipath cues such as AoA and relative time-of-flight while providing a stable and physically meaningful learning objective from sparse CSI.

\Raf{We evaluate \sysname{} as a high-fidelity spatial planning tool using CSI collected at various positions using RT-AC86U commodity WiFi routers with four antennas and 50 OFDM subcarriers (20MHz bandwidth channel) in the 5 GHz band.} Experiments span indoor environments with different room scales and furniture configurations, enabling evaluation under diverse multipath conditions, including both \gls{los} and \gls{nlos} regions.
 We compare \sysname{} against existing neural channel modeling baselines, including NeRF$^{2}$ \cite{zhao2023nerf2}, NeWRF \cite{lu2024deep}, VAE \cite{kingma2013auto}, and GAN \cite{goodfellow2014generative}. \rev{\sysname{} estimates complex-valued \gls{csi} matrix at held-out receiver locations within the same environment with high fidelity}, achieving a median SNR of 5.30~dB, \textbf{outperforming state-of-the-art baselines by 3.0--6.7~dB (131--368\% improvement)}, while using only 4 samples per cubic feet of training data. Beyond channel estimation accuracy, we demonstrate the \saif{offline} actionability of the estimated \gls{csi} and its preserved phase coherence through three representative applications: beamforming gain evaluation, spatial signal coverage mapping for wireless-aware path planning, and \gls{los} \gls{aoa} estimation using the classical MUSIC algorithm~\cite{kotaru2015spotfi, schmidt1986multiple}. Importantly, all evaluations are performed using standard signal processing methods directly on the predicted CSI, rather than training downstream estimators. This ensures that observed performance reflects true physical fidelity and learned multipath structure rather than task-specific overfitting. %

\saif{As such, our contributions are as follows}:
\begin{itemize}
    \item We introduce \sysname{}, a neural field framework for wireless channel modeling that directly incorporates the measurement limits of commodity devices as inductive constraints, enabling structured learning of complex-valued CSI from sparse data.

    \item We propose a cone-based wave sampling formulation that models hardware-imposed angular resolution and ambiguity, limiting the learned representation to multipath components that commodity antenna arrays can actually resolve.

    \item We incorporate a multi-resolution implicit scene representation that jointly captures large-scale environmental structure and fine-grained scattering effects, allowing accurate channel prediction at unseen receiver locations \rev{within the same environment.}

    \item We present a rendering-based training pipeline with relative phase learning objective that enables stable learning from noisy commodity CSI, and show that the learned channel representations can be directly used for downstream tasks such as beamforming and \gls{aoa} estimation in real indoor environments.

\end{itemize}

\section{Background and Related Work}

\subsection{Wireless Channel Estimation} %
Wireless channel modeling has been a central topic in wireless communication \cite{zheng2020fast, goldsmith2005wireless}, with a more recent trend on using data-driven models \cite{wang2019ul}. The main paradigm is to use machine learning to learn complex channel distributions and model the channel input-output relationships as a conditional probability distribution \cite{walfisch1988theoretical}, generative adversarial networks (GANs) \cite{ye2020deep}, or variational auto-encoders (VAEs) \cite{dorner2020wgan}.  However, these methods have seen limited practical adoption due to their reliance on large training datasets and their limited ability to generalize to scenarios not observed during training. Another category of wireless channel modeling approach focuses on wireless simulation and ray-tracing \cite{ji2001efficient} with more recent techniques leveraging deep learning for optimizing the ray tracing \cite{hoydis2023sionna}. However, they rely on accurate 3D model of the environment and large training data.

\subsection{Neural Radiance Fields for Wireless} %
Recent advances in neural radiance fields have transformed visual computing by representing physical scene properties as continuous functions in space. The key strength of this approach is the ability to construct spatial models from sparse observations and predict signal characteristics at unseen locations \rev{within a scene}. Viewed through the same lens as the original definition of physical fields in physics \cite{feynman1965feynman}, a wireless propagation environment can also be parameterized as a field that maps spatial coordinates to electromagnetic quantities such as complex attenuation, phase, or frequency response. Motivated by this idea, recent work has proposed \gls{nerf}-inspired models to synthesize wireless channels.  WiNeRT \cite{orekondy2023winert} and NeWRF \cite{lu2024deep}, for example, aim to reconstruct complete Channel Impulse Response or Channel Frequency Response fields that capture multipath amplitude, phase, and delay. However, these models rely exclusively on synthetic data because they require per-path information such as Angle of Arrival and Time of Flight for training or evaluation, and such annotations are not practical to acquire from real-world measurements.

Another line of work focuses on synthesizing RF measurements for downstream tasks. Examples include NeRF² \cite{zhao2023nerf2} and RF-Canvas \cite{chen2024rfcanvas}, which primarily estimate RSSI or spatial spectrum rather than full complex propagation behavior and therefore do not demonstrate the phase accuracy required for coherent MIMO processing. Furthermore, these methods rely on unrealistic assumptions such as extremely dense measurements or a known coarse geometry of the scene. These assumptions reflect the fundamentally ill-posed nature of wireless channel reconstruction, where many different scattering configurations can produce similar sparse CSI. Instead, our method \rev{incorporates} the intrinsic measurement limits of commercial WiFi hardware as constraints in the model, enabling physically observable structure to guide learning and allowing complex wireless channels to be estimated from sparse, noisy experimental data.

\begin{figure*}[t]
    \centering
    \includegraphics[width=0.99\textwidth]{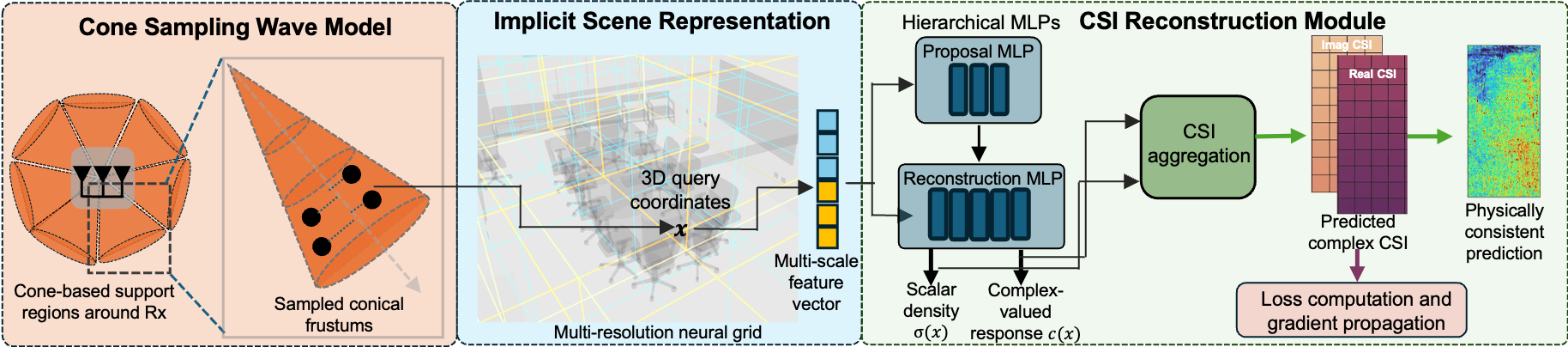} 
    \caption{\rev{Given sparse spatial \gls{csi} measurements, \sysname{} reconstructs a continuous complex channel field through three modules: (1) \wavemodel, which defines cone-based support regions and sampled conical frustums around the receiver; (2) \worldmodel, which encodes the (x,y,z) samples using a multi-resolution hash grid; and (3) \rendering, which predicts and aggregates complex \gls{csi}.}}
    
    \label{fig:volumetric_grid_encoding}
\end{figure*}

\vspace{0.3em}
\noindent \textbf{Architectural Limits of \gls{nerf}-based Wireless Models:}
These limitations point to two recurring architectural mismatches in prior \gls{nerf}-based wireless models: how propagation is sampled and how scene structure is represented across scales. The existing papers primarily apply the classical \gls{nerf} architecture directly to the wireless domain, suffering from several architectural limitations that our method addresses:
 \begin{itemize}[leftmargin=*]
   \item \textbf{Inefficient Sampling:} \gls{nerf} assumes dense ray sampling along infinitesimally thin rays, which works for a pinhole camera where each pixel corresponds to a unique direction and megapixel images yield millions of angular samples \cite{mildenhall2021nerf}. In wireless, the receiver antenna array is the sensor, and its \gls{aoa} resolution is fundamentally limited by the number of elements, aperture size, and SNR \cite{van2002optimum, paulraj2003introduction}. Therefore, directly porting \gls{nerf} produces two inefficient extremes: dense ray casting with huge training sets as in \gls{nerf}$^2$ \cite{zhao2023nerf2}, which is wasteful because most rays do not correspond to physical paths, or sparse ray casting guided only by a few AoAs as in NeWRF \cite{lu2024deep}, which is highly susceptible to aliasing and hardware-limited angular resolution, resulting in inaccurate field reconstruction when real-world data is used.
   \item \textbf{Scale Ambiguity:} Classical \gls{nerf} uses a single MLP to represent the entire scene, and most NeRF-based wireless methods inherit this same architecture. However, wireless propagation involves both large, dominant reflectors (such as walls) and fine, localized scattering features (such as furniture edges). These different scattering mechanisms operate at different spatial scales and impose very different structure on the signal. Prior works such as WiNeRT \cite{orekondy2023winert} and RF-Canvas \cite{chen2024rfcanvas} attempt to cope with this by injecting explicit geometry priors, but this is often impractical because it requires either perfect 3D scene models from simulation or additional sensing modalities that are not available in real deployments. Without such priors, a single MLP tends to oversmooth (spectral-bias) and fails to capture the sharp multipath variations.
 \end{itemize}

Recent \gls{nerf} progress in computer vision offers architectural ideas that can help address these structural limitations for wireless: (i) Anti-aliasing: Mip-\gls{nerf} \cite{Barron_2021_ICCV} replaces infinitely thin rays with cones and integrates features over their finite footprint to explicitly suppress aliasing artifacts. (ii) Multi-resolution encoding: Instant-NGP \cite{muller2022instant} introduces a multi-resolution hash-grid positional encoding so that coarse cells capture large structures while fine cells capture small details. More recently, Zip-\gls{nerf} \cite{Barron_2023_ICCV} combines both concepts, using cone footprints to select the appropriate hash-grid scale at each query location. However, Zip-\gls{nerf} assumes a dense camera pixel grid that preserves angular information, which wireless systems do not provide because RF measurements are sparse, array-limited, and do not directly yield pixel-level direction vectors. Inspired by Zip-\gls{nerf}, we therefore propose a new wave-and-world field formulation that addresses the inefficient sampling and scale ambiguity discussed above, but repurposes these ideas for wireless by explicitly embedding hardware-dependent observability into the representation.

\section{System Design}
\label{sec:methodology}
\sysname{} introduces a neural field framework that replaces the idealized line-ray abstraction in classical NeRF formulations with a cone-based volumetric wave model. \saif{This design explicitly reflects the finite angular resolution of commodity antenna arrays, which fundamentally limits how precisely incoming propagation directions can be distinguished \cite{van2002optimum}. As a result, commodity WiFi devices observe propagation only up to their finite angular resolvability, collapsing multiple incoming directions into indistinguishable angular sectors rather than infinitesimally precise rays.} In contrast, prior NeRF-based wireless models \cite{zhao2023nerf2, lu2024deep} assume arbitrarily fine angular resolution by representing propagation with infinitesimally thin rays. This abstraction is mismatched with real antenna observability and leads to \gls{aoa} aliasing: thin rays represent directions that are indistinguishable at the hardware level and therefore unlikely to consistently intersect resolvable multipath structure in the scene. \rev{This mismatch often results in physically implausible channel estimates and weaker spatial prediction at unseen receiver locations within the same environment.}

An overview of the end-to-end workflow of \sysname{} is shown in Figure.~\ref{fig:volumetric_grid_encoding}. Without loss of generality, we focus on a \gls{ula}, though the formulation naturally extends to other array geometries. \saif{The framework consists of three tightly coupled components: (1) a \textbf{\wavemodel}, (2) a \textbf{\worldmodel}, and (3) a \textbf{\rendering} module.} We detail each component in the following subsections.

\begin{figure}[t]
    \centering
    \includegraphics[width=\linewidth]{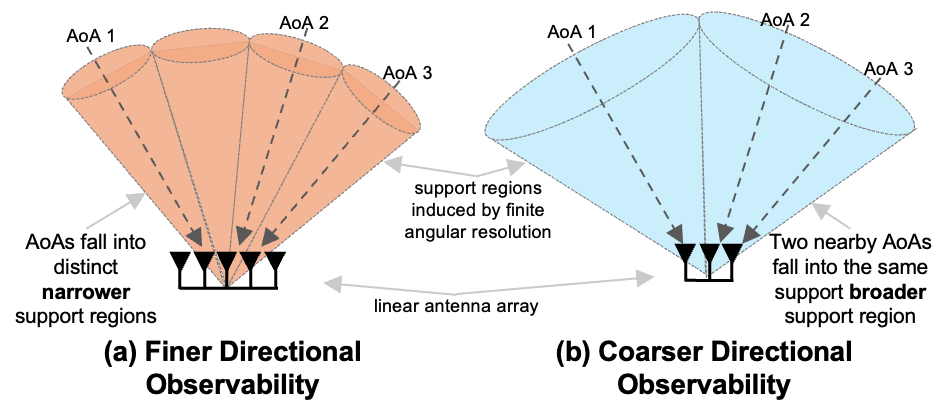}
    \caption{\rev{Array-dependent angular support in \sysname{}. The figure schematically compares two \gls{ula} configurations with different numbers of antennas: (a) higher-element \gls{ula} with finer directional observability and (b) lower-element \gls{ula} with coarser directional observability. \sysname{} adjusts cone-based support width accordingly. With finer observability, three example \gls{aoa}s occupy distinct conical support regions; with coarser observability, the broader support regions are grouped into two nearby \gls{aoa}s.}}
    \label{fig:wedge_ray}
\end{figure}

\subsection{\saif{\wavemodel}}
\label{subsec:wave_model}

The goal of \sysname{}'s \saif{wave model is to incorporate antenna-induced observability limits as inductive biases that constrain what the neural field can represent.} Rather than attempting to compensate for hardware imperfections through post-processing or learning arbitrary corrections, the model is explicitly restricted to represent only those angular structures that are distinguishable given the finite resolution of the antenna array. By construction, multipath components that are closer in angle than the array’s resolvable limit are treated as indistinguishable and are not artificially separated during learning. This alignment between the sampling model and the measurement process encourages physically consistent channel representations that remain usable for downstream tasks such as beamforming and \gls{aoa} estimation.

\vspace{.3em}\noindent\textbf{Cone Geometry:} In \sysname{}, \saif{angular sampling is performed using volumetric cones rather than infinitesimally thin rays. \rev{Importantly, these cones are not intended to approximate the full beamforming process, directional radiation pattern, or exact array response function. Instead,} they represent the set of arrival directions that are indistinguishable under the antenna array’s finite angular resolution}. Each cone is parameterized by a central direction vector $\mathbf{u}$ and two angular widths: the azimuthal width $\alpha_{\mathrm{az}}$ in the horizontal plane and the elevation width $\alpha_{\mathrm{el}}$ in the vertical plane.

For a cone centered at the antenna array, the cross-sectional radii at distance $d$ are given by
\begin{equation}
r_{\mathrm{az}}(d) = d \tan\left(\frac{\alpha_{\mathrm{az}}}{2}\right), \quad
r_{\mathrm{el}}(d) = d \tan\left(\frac{\alpha_{\mathrm{el}}}{2}\right).
\end{equation}

This formulation defines an anisotropic (elliptical) cone whose shape reflects direction-dependent angular resolution imposed by the array geometry. The linear expansion with distance preserves constant angular uncertainty, consistent with array resolution being determined by aperture and antenna element placement. \rev{As a result, the cone geometry serves as a measurement-constrained inductive bias that regularizes the inverse problem under \gls{aoa} aliasing, rather than learned smoothing or a claim of full electromagnetic fidelity.}

\rev{More generally, \sysname{} sets cone width according to the directional observability of the receiving array: arrays that can resolve directions more finely are assigned narrower angular support regions, while arrays with coarser directional observability are assigned broader ones. Figure~\ref{fig:wedge_ray} schematically illustrates this idea using three example AoAs. In the finer-observability case, the three AoAs occupy distinct support regions and therefore remain distinguishable at the level of the measurement model. In the coarser-observability case, two nearby AoAs fall within the same broader support region, indicating that they are not separately resolved by the array under the assumed observability. The figure is intended as a conceptual illustration of support-region width, not as a depiction of beamforming patterns or exact array response.}

\begin{figure}[t]
    \centering
    \includegraphics[width=\linewidth]{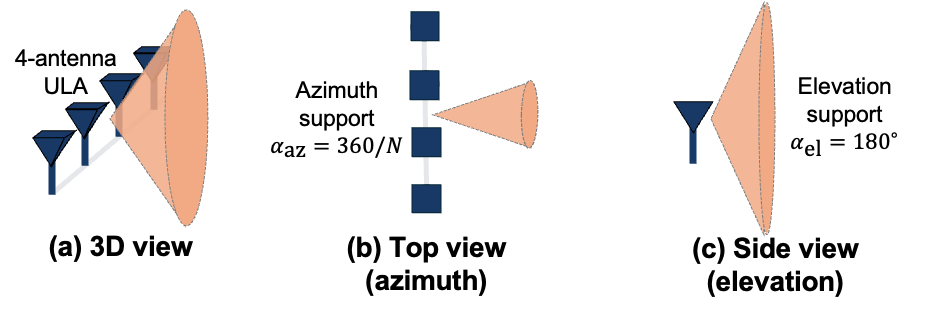}
    \caption{Angular support region for a 4-antenna \gls{ula}.}
    \label{fig:ula_support_region}
\end{figure}

As one concrete example, consider a \gls{ula} with four antenna elements spaced at half-wavelength intervals. Such an array provides directional discrimination primarily in azimuth, while exhibiting strong ambiguity in elevation due to the absence of vertically separated elements. To reflect this asymmetry, \sysname{} assigns the azimuthal cone width as $\alpha_{\mathrm{az}} = \frac{360}{N}^\circ$, where $N$ denotes the number of angular partitions around the receiver, and sets $\alpha_{\mathrm{el}} = 180^\circ$ to capture near-complete elevation ambiguity. \rev{Figure~\ref{fig:ula_support_region} illustrates the resulting anisotropic support region used in \sysname{} for a 4-antenna \gls{ula}, shown in 3D, top, and side views. The figure depicts one representative support region; in practice, \sysname{} defines multiple such regions around the receiver by partitioning the azimuth plane into discrete sectors.} \rev{In our implementation, this observability model is instantiated using 12 azimuthal sectors of $30^\circ$ each, which are used consistently during both training and inference.}

Each cone is discretized into $K$ segments, or conical frustums, along the propagation path to enable volumetric integration. Let $\mathbf{o}$ denote the ray origin (antenna position), $\mathbf{u}$ the unit direction vector, and $d_k$ the midpoint distance of the $k$-th segment from the origin. The center of each frustum is given by $\boldsymbol{\mu}_k = \mathbf{o} + d_k \mathbf{u}$. To represent the spatial extent of each frustum, \sysname{} employs the supersampling strategy from Zip-NeRF~\cite{zipnerf}. In this formulation, each sample within a frustum is modeled as a Gaussian distribution parameterized by its mean and variance. The variance reflects the local volume of the frustum as smaller frustums correspond to narrower Gaussians. This allows the representation to adapt continuously to the geometric scale of the segment. By sampling multiple Gaussian-distributed points within each frustum, the model captures both the mean and variance (first and second spatial moments) of the enclosed volume, ensuring that the aggregated features accurately approximate its volumetric contribution. This approach provides a differentiable alternative to explicit Gaussian volume modeling, as the sampling pattern and segment length $\Delta d_k = d_{k+1} - d_k$ implicitly encode the axial extent of each segment without requiring a separate axial variance term.%

\subsection{\saif{\worldmodel}}
\label{subsec:world_model}

The cone-based wave model in \sysname{} induces spatial queries with varying footprints and scales. Narrow conical frustums emphasize localized spatial variation, while wider frustums aggregate information over larger regions. Moreover, the spatial extent of each frustum increases with distance from the antenna array, reflecting the fact that array-induced angular uncertainty corresponds to larger spatial ambiguity in the far field. As a result, a single frustum may simultaneously span coarse structures (e.g., walls or floors) and finer scattering features (e.g., edges or furniture). Representing such mixed-scale interactions using a single fixed-scale representation would blur fine spatial variation into coarse background structure and limit expressivity.

To address this, \sysname{} employs a multi-resolution implicit scene representation that supports spatial feature encoding across scales. Rather than explicitly reconstructing geometry or material properties, this representation learns a latent, spatially indexed feature field that is queried exclusively through cone-based sampling. This design aligns with the measurement-driven nature of the system and avoids imposing external geometric or visual priors. Figure~\ref{fig:concept} illustrates this distinction: unlike classical line-ray formulations that rely on a single-scale MLP, \sysname{} uses a multi-resolution hash grid~\cite{instantNGP} to efficiently support coarse-to-fine spatial variation while maintaining memory efficiency and training stability.

\vspace{.3em}\noindent\textbf{Multi-Resolution Hash Grid Architecture:}
The implicit scene representation maps each 3D location $\mathbf{x} = (x, y, z)$ to a feature vector constructed from $L$ resolution levels. At each level $l \in [1, L]$, space is discretized into a regular grid, with features stored in a fixed-size hash table to enable high effective resolution without incurring the memory cost of dense voxel grids. The coarse grid levels capture room-scale structures while fine levels resolve small scatterers and detailed geometry.

For each resolution level $l$, the feature embedding at location $\mathbf{x}$ is computed via trilinear interpolation over the eight nearest grid vertices:
\begin{equation}
f^{(l)}(\mathbf{x}) = \sum_{v \in \mathcal{N}(\mathbf{x}, l)} w_v(\mathbf{x}) \cdot \mathbf{g}^{(l)}(h(v)),
\end{equation}
where $\mathcal{N}(\mathbf{x}, l)$ denotes the neighboring grid vertices at level $l$, $w_v(\mathbf{x})$ are the trilinear interpolation weights, $h(\cdot)$ is the spatial hash function, and $\mathbf{g}^{(l)}(\cdot)$ are trainable feature vectors stored in the hash table. \rev{Importantly, this interpolation is applied only to latent multi-scale scene features, not directly to absolute RF phase or to a full electromagnetic field representation. This allows the model to preserve coarse-to-fine spatial continuity in the learned representation without claiming smoothness of the underlying electromagnetic field itself.} Features from all resolution levels are concatenated to form a multi-scale embedding:
\begin{equation}
\mathbf{f}(\mathbf{x}) = \bigoplus_{l=1}^{L} \mathbf{f}^{(l)}(\mathbf{x}),
\end{equation}
which is then passed to downstream neural networks for channel synthesis.

This representation provides three key benefits:
(i) \emph{Compactness}: Hash-based table reuses a small pool of trainable features instead of storing full voxel grids at each resolution, greatly reducing memory.
(ii) \emph{Multi-scale support}: Coarse grid levels capture large structures such as walls and ceilings, while fine levels resolve small scatterers like furniture and clutter.
(iii) \emph{Continuity}: Trilinear interpolation ensures smooth feature variation \rev{in the latent scene features} with respect to spatial location, which is important for stable volumetric aggregation \rev{without directly imposing smoothness on absolute RF phase.}

\vspace{.3em}\noindent\textbf{\sysname{} Neural Network Architecture: }
Conditioned on the multi-resolution scene features, \sysname{} predicts intermediate latent quantities that govern how samples along each cone contribute to channel synthesis. Specifically, for each sampled location $\mathbf{x}$, the network estimates:
(1) a scalar density $\sigma(\mathbf{x})$, which acts as a soft weighting function that modulates the relative contribution of the location during volumetric aggregation along the cone, and
(2) a complex-valued response $\mathbf{c}(\mathbf{x}) \in \mathbb{C}$, which encodes amplitude scaling and phase rotation applied to the propagating signal.
\saif{These quantities are not interpreted as explicit physical material properties or scattering parameters. Instead, they are learned latent variables whose sole role is to support accurate reconstruction of observed \gls{csi} under the cone-based sampling and aggregation process.}

To improve computational efficiency, \sysname{} adopts a hierarchical two-stage architecture inspired by Zip-NeRF~\cite{zipnerf}. A lightweight proposal network first evaluates coarse samples along each conical ray to estimate a rough volumetric density distribution, identifying spatial regions with non-negligible contributions to the received signal. The input to the proposal network is the multi-resolution hash grid encoding corresponding to each cone frustum sample, enabling efficient assessment of where meaningful interactions are likely to occur.

Guided by the proposal network’s weights, a second network then processes a refined set of samples, focusing computational resources on regions that contribute most strongly to the channel. This refined network predicts both the scalar density $\sigma(\mathbf{x})$ and the complex-valued response $\mathbf{c}(\mathbf{x})$ used for final channel synthesis. To ensure consistency between stages, the proposal network is trained with auxiliary consistency losses that encourage its density estimates to align with the refined network’s outputs, preventing divergence while significantly reducing the number of expensive network evaluations.

\subsection{\saif{\rendering}}
\label{subsec:rendering}

\saif{\sysname{} reconstructs channel state information by aggregating learned complex responses along cone-shaped sampling volumes defined by the wave model. This module combines cone-based sampling with the implicit scene representation to synthesize \gls{csi} in a manner consistent with the angular resolution and ambiguity of the antenna array.} A key challenge in this aggregation is aligning the continuous spatial support of conical frustums with the discrete multi-resolution hash grid used to encode the scene. \sysname{} addresses this using a level-of-detail (LOD) weighting strategy adapted from Zip-NeRF~\cite{zipnerf}, where the spatial extent of each frustum determines how features from different grid resolutions contribute to the final response.

\begin{figure}[t]
    \centering
    \includegraphics[width=\linewidth]{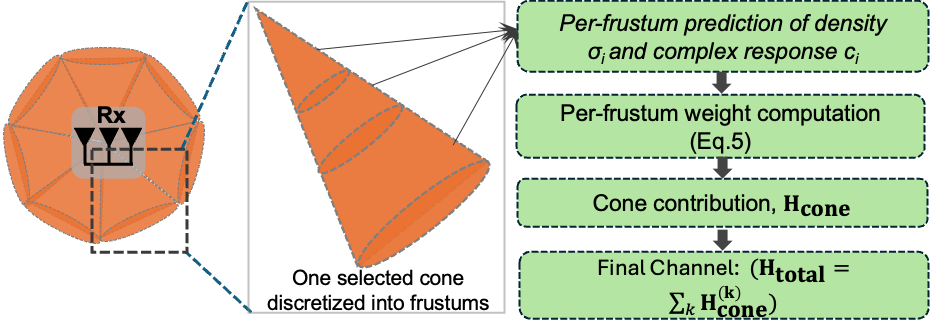} 
\caption{\rev{The full pipeline of CSI estimation via per-frustum prediction and aggregation.}}

    \label{fig:volumetric_rendering_loss}
\end{figure}

The channel reconstruction process begins by partitioning the angular space around each receiver into $P$ non-overlapping bimodal wedges, where each wedge corresponds to a cone of arrival directions defined by the array’s intrinsic angular resolution. Within each cone, contributions are accumulated along the propagation direction by discretizing the cone into $N$ conical frustums. For a given cone, the complex channel contribution is computed as a weighted sum over frustums:
\begin{equation}
\mathbf{H}_{\text{cone}} = \sum_{i=1}^{N} w_i \, \mathbf{c}_i ,
\end{equation}
\saif{where $\mathbf{c}_i$ denotes the complex-valued response predicted at frustum $i$, and $w_i$ is a scalar weight that modulates the relative contribution of that location during aggregation.} The weights are computed using an exponential attenuation formulation:
\begin{equation} \label{eq:weights}
w_i = T_i \cdot \alpha_i, \quad 
\alpha_i = 1 - \exp(-\sigma_i \delta_i), \quad
T_i = \exp\Big(-\sum_{j=1}^{i-1} \sigma_j \delta_j\Big),
\end{equation}
where $\sigma_i$ is the learned density at frustum $i$, $\delta_i$ is its axial length, $\alpha_i$ controls the local contribution strength, and $T_i$ accumulates the residual contribution from preceding frustums along the cone. These quantities are treated as learned aggregation variables rather than explicit physical interaction probabilities. Figure ~\ref{fig:volumetric_rendering_loss} depicts the channel reconstruction process. The full channel response at a receiver is obtained by summing contributions from all cones:
\begin{equation}
\mathbf{H}_{\text{total}} = \sum_{k=1}^{P} \mathbf{H}_{\text{cone}}^{(k)},
\end{equation}
where $\mathbf{H}_{\text{cone}}^{(k)}$ denotes the aggregated response from the $k$-th cone. To support multi-antenna and multi-subcarrier CSI, the network predicts complex responses with dimensions $N_{\text{rx}} \times N_{\text{sc}}$, enabling simultaneous reconstruction across spatial and frequency domains within a unified framework.

\saif{It is worth noting that the entire reconstruction pipeline, that progresses from cone initialization and frustum sampling to weighted aggregation and CSI synthesis, is fully differentiable. This allows end-to-end optimization of both the implicit scene representation and the neural networks using gradient-based learning.}

\subsection{\saif{Training Objective}}
\label{subsec:training}

\saif{\sysname{} is trained by aligning synthesized channel responses from the forward rendering pipeline with measured CSI through end-to-end differentiable optimization. Rather than performing inverse rendering or recovering explicit physical scene parameters, training constrains the latent neural representation so that its forward-rendered CSI matches observations collected by commodity hardware. Mini-batches of receiver locations are sampled from the training set, and the corresponding measured CSI matrices serve as supervision. Gradients propagate from the CSI reconstruction loss through the channel aggregation, implicit scene representation, and cone-based sampling process.}

A central challenge when training on experimental CSI is that the absolute phase measured by commodity devices is unreliable. Hardware-induced effects such as \gls{cfo}, \gls{sfo}, and packet detection delay introduce packet-dependent global phase drifts that are unrelated to propagation \cite{kotaru2015spotfi}. As a result, directly regressing absolute phase forces the model to explain device-specific artifacts rather than multipath structure.

\saif{Following common practice in WiFi CSI processing, we first apply standard packet-level phase sanitization to mitigate global \gls{cfo} and \gls{sfo} components that are shared across antennas and subcarriers within each snapshot \cite{kotaru2015spotfi, liu2021fire}. However, residual phase offsets and snapshot-to-snapshot drift remain unavoidable due to hardware instability and measurement noise, making absolute phase values inconsistent across packets \cite{tan2006adaptive, liu2021fire}.}

\saif{To ensure that learning focuses on stable and physically informative structure, we propose a relative phase loss that operates on \rev{unwrapped,} antenna-referenced phase differences rather than absolute phase. This removes remaining snapshot-level global phase offsets while preserving spatial phase differences across antennas and subcarriers, which encode relative angle-of-arrival and delay information under the array geometry. \cite{goldsmith2005wireless, music}.} For each CSI snapshot, we normalize the phase by subtracting the phase of a reference antenna element from all antenna channels. This removes snapshot-level global phase drift while preserving spatial phase differences across antennas and subcarriers, which encode relative \gls{aoa} and delay information. Both the predicted channel $\hat{\mathbf{H}}$ and the measured channel $\mathbf{H}$ are normalized in this manner prior to loss evaluation.

The training objective minimizes the normalized mean squared error between the relative-phase CSI representations. Let $\mathbf{H} \in \mathbb{C}^{N_r \times N_f}$ denote the measured CSI at a receiver with $N_r$ antennas and $N_f$ subcarriers.
We define the relative-phase CSI $\mathbf{H}_{\text{rel}}$ by normalizing each CSI entry with respect to a reference antenna--subcarrier pair:
\begin{equation}
\mathcal{L}_{\text{rel-phase}} =
\frac{
\left\lVert 
\hat{\mathbf{H}}_{\text{rel}} 
- 
\mathbf{H}_{\text{rel}} 
\right\rVert^2
}{
\left\lVert 
\mathbf{H}_{\text{rel}} 
\right\rVert^2
}
\end{equation}

\saif{This loss does not aim to recover absolute phase or correct hardware impairments. Instead, it encourages the learned neural field to reproduce the relative phase structure \rev{ supported by the measurements under the cone-based sampling and rendering process, rather than arbitrary sub-wavelength absolute field oscillations}, enabling accurate \gls{csi} reconstruction for downstream multi-antenna tasks.}

\section{Implementation}

\subsection{Data Collection}

\begin{figure}[t]
    \centering
    \begin{subfigure}[t]{0.49\linewidth}
        \centering
        \includegraphics[width=\linewidth]{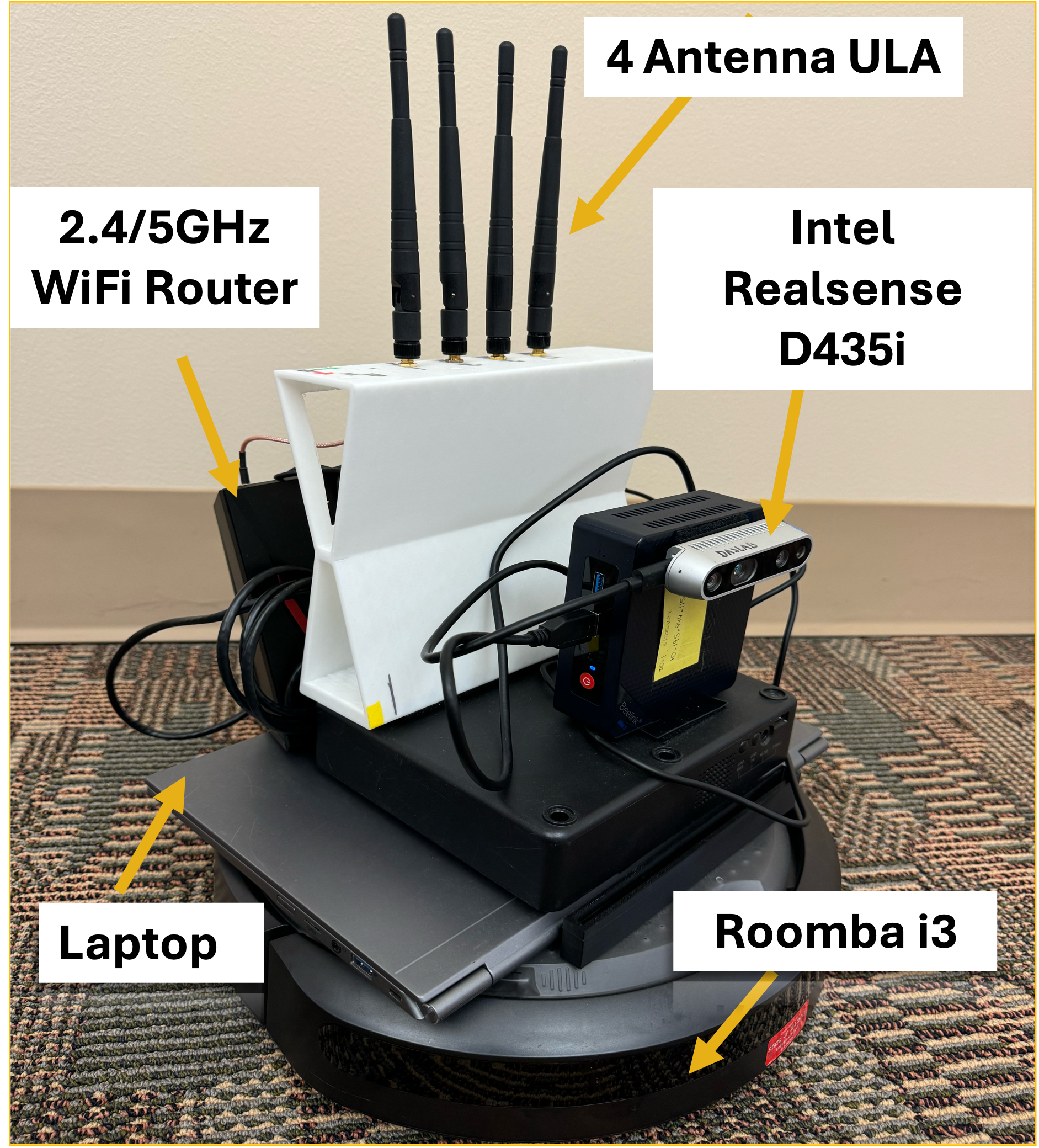}
        \caption{Mobile Platform for\\data collection}
        \label{fig:data_setup}
    \end{subfigure}
    \hfill
    \begin{subfigure}[t]{0.49\linewidth}
        \centering
        \includegraphics[width=\linewidth]{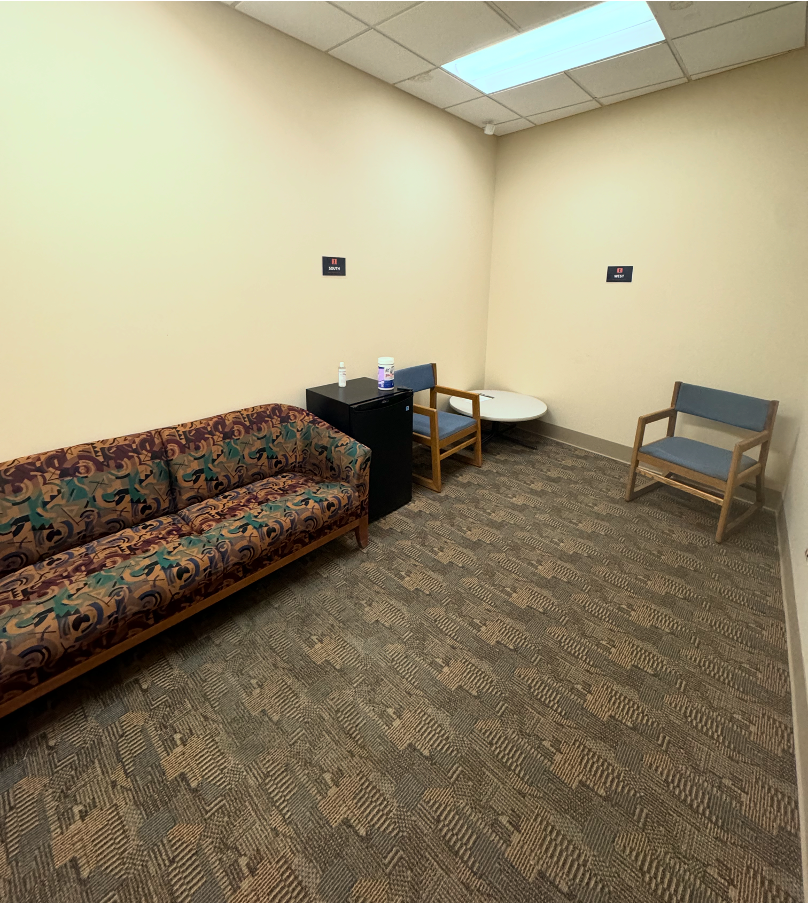}
        \caption{Meditation Room\\(4m $\times$ 2.25m)}%
        \label{fig:room1}
    \end{subfigure}
    \hfill
    \begin{subfigure}[t]{0.49\linewidth}
        \centering
        \includegraphics[height = 4.6cm, width=\linewidth]{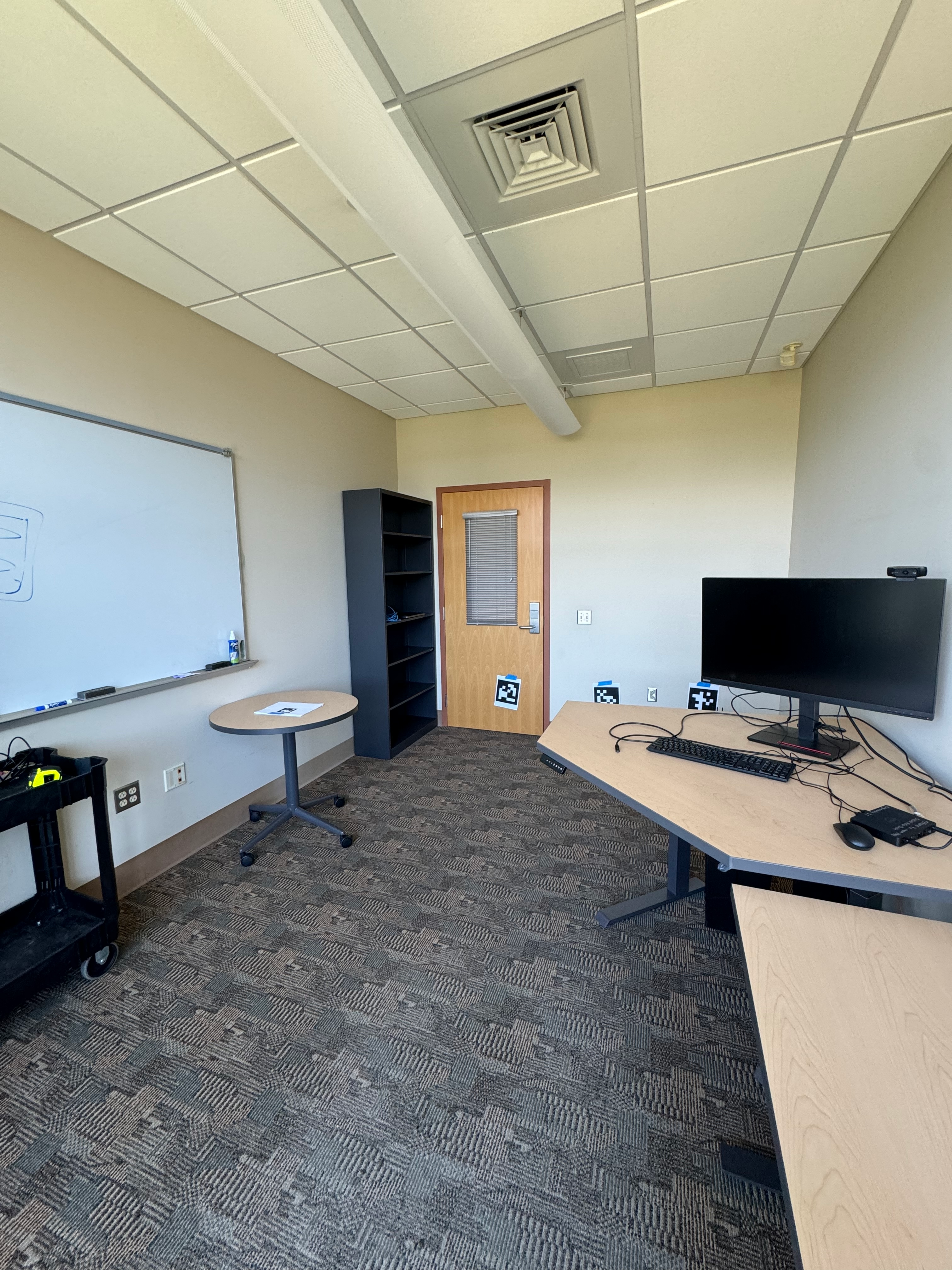}
        \caption{Office Space\\(5m $\times$ 3m )}%
        \label{fig:room2}
    \end{subfigure}
    \hfill
    \begin{subfigure}[t]{0.49\linewidth}
        \centering
        \includegraphics[width=\linewidth]{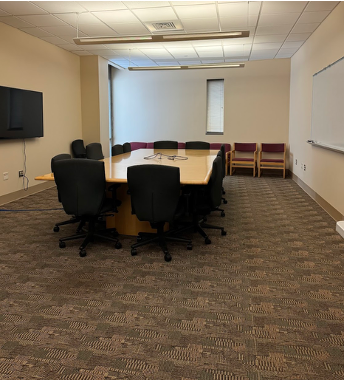}
        \caption{Conference Room\\(12m $\times$ 6m )}%
        \label{fig:room3}
    \end{subfigure}

    \caption{\rev{(a) Mobile platform used for CSI data collection with a 4-antenna WiFi router, Intel Realsense depth camera, and laptop mounted on a Roomba. (b-d) WiNeRF experimental environments. }}

    \label{fig:simulation_env_models}
\end{figure}

\noindent\textbf{Mobile Platform:} As shown in Figure~\ref{fig:data_setup}, we employ a mobile platform based on iROBOT Create 3 to collect the spatial CSI data. The robot is equipped with an ASUS RT-AX86U router, configured as a receiver (Rx). The router's firmware was modified using Nexmon CSI~\cite{nexmon} tool to enable CSI extraction, and WiROS~\cite{wiros} framework is used to collect CSI and RSSI. We also use an Intel RealSense depth D435i camera for simultaneous localization and mapping (SLAM) based on ORBSLAM3 \cite{orbslam3} to track the 6-DoF pose of the robot. The robot was moved manually throughout the space using random trajectories. 

\vspace{.3em}\noindent\textbf{Wireless Devices:} We collected CSI data on a 20 MHz channel in the 5 GHz band (Channel 157), resulting in 50 active subcarriers captured by 4 Rx antennas from a single spatial stream. To address the inherent phase noise in commodity WiFi hardware, we implemented a two-step calibration process. First, we modified the firmware to disable automatic receive-chain switching. Second, we performed a dynamic phase-compensation procedure \cite{wiros} prior to data collection to ensure that the CSI matrices retain a consistent and correct phase. Maintaining phase coherence across antennas and subcarriers is essential because the relative phase encodes the channel’s underlying multipath structure. Without proper alignment, the signatures of individual paths become distorted, which in turn affects any downstream processing that relies on stable spatial or frequency characteristics. With the compensated phase, the resulting CSI accurately reflects the true multipath profile of the environment.

\vspace{.3em}\noindent\textbf{Experimental Environments:} We collected data in three distinct indoor environments: a Meditation Room (\textit{4\,m $\times$ 2.25\,m $\times$ 2.4\,m}), an Office Space (\textit{5\,m $\times$ 3\,m $\times$ 2.7\,m}), and a Conference Room (\textit{12\,m $\times$ 6\,m $\times$ 2.4\,m}), resulting in a total of 30k synchronized CSI and pose samples, with an average data sampling density of 4 samples per~ft$^3$ (equivalent of 150 samples per meter$^3$). All environments contain various furniture configurations and exhibit rich multipath propagation conditions. The Conference Room dataset includes numerous non-line-of-sight (NLoS) cases caused by a large table positioned at the same height as the robotic platform, which blocks the line-of-sight (LoS) path between the transmitter and receiver across most locations. The Office Space, on the other hand, contains several metal cabinets that enhance multipath propagation and increase signal reflections. A snapshot of all environments is shown in Figs.~\ref{fig:room1}, \ref{fig:room2}, and \ref{fig:room3}.

\subsection{Model Training and Configuration}

We implement \sysname{} in PyTorch (v2.3) with CUDA~12 and employ the AdamW optimizer for end-to-end training. All experiments are conducted on \rev{our server} equipped with two NVIDIA RTX~A6000 GPUs. Training uses a batch size of 128, an initial learning rate of $10^{-3}$ with cosine decay to $10^{-4}$, and a maximum of 50{,}000 epochs. \rev{For each environment, \sysname{} is trained separately to predict} the full complex CSI matrix for 4 receive antennas across 50 OFDM subcarriers, \rev{and the current system does not include a cross-environment transfer or rapid adaptation mechanism.} Evaluations are conducted using 5-fold cross-validation, where each fold uses an 80\%/20\% split between training and testing samples. \rev{The per-fold wall-clock training time is approximately 15--30 minutes on our RTX~A6000 server, depending on the dataset split and scene complexity.}

Overfitting is mitigated by the broad diversity of CSI measurements across spatial positions, LoS/NLoS regions, and multipath patterns induced by metal reflectors. Combined with the strong inductive biases of our NeRF-style architecture \cite{nerf, instantNGP, zipnerf} (e.g., volumetric density fields and hash-grid feature sharing), the model is encouraged to learn physically consistent propagation behavior instead of memorizing individual CSI samples.

\rev{WiNeRF uses one proposal network and one reconstruction network. The proposal network employs a multiresolution hash grid with a base resolution of $16^3$ and a finest resolution of $256^3$, corresponding to 5 resolution levels, and contains 0.86M learnable parameters. This proposal stage efficiently identifies regions of significant multipath energy. The reconstruction network uses a larger multiresolution hash grid with the same base resolution of $16^3$ and a finest resolution of $2048^3$, corresponding to 8 resolution levels, and contains 1.87M learnable parameters. Across both networks, WiNeRF contains 2.73M learnable parameters in total.}

\rev{The reconstruction network processes hash-encoded features through two multi-layer perceptrons (MLPs). A density MLP (2 layers, 64 units per layer) outputs volumetric density along with a 64-dimensional bottleneck vector. A view-conditioned response branch (2 layers, 256 units per layer) combines this bottleneck with directionally encoded viewing cones to predict the complex-valued response for all subcarriers and receiver antennas. Under our setup with 4 receive antennas, 1 transmit antenna, and 50 subcarriers, this branch predicts the real and imaginary parts of the full complex CSI tensor. The final complex CSI is computed via volumetric integration, where the predicted responses are weighted by the volumetric density and integrated along each cone to synthesize the channel response.}

\vspace{.3em}\noindent \textbf{Cone Quantization:}
To operationalize the observability-driven cone model in our 4-antenna ULA setting, we divide the azimuth plane into 12 equal $30^\circ$ sectors. During both training and inference, we emit \textit{bimodal cone-wedge} support regions from the receiver for each discretized azimuth bin. Each bimodal cone-wedge has two properties: (i) its elevation span covers the entire elevation range, since our receiver array does not provide elevation resolution, and (ii) its azimuth span is restricted to the corresponding $30^\circ$ bin, reflecting the angular discretization used in our model. This implementation directly mirrors the finite angular observability assumed in the wave model, allowing the representation to remain physically meaningful while staying aligned with the hardware's actual directional resolution.

\captionsetup[subfigure]{justification=centering,singlelinecheck=false}

\begin{figure*}[t]
  \hspace*{-0.06\textwidth} 
  \centering
  \begin{subfigure}[t]{0.26\textwidth}
    \centering
    \includegraphics[width=\textwidth]{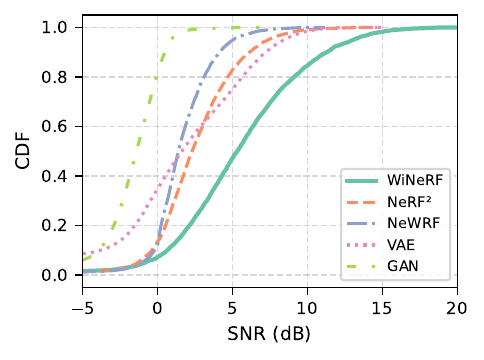}
    \caption{Prediction SNR \\ (Higher is better)}
    \label{fig:snr_cdf}
  \end{subfigure}
  \begin{subfigure}[t]{0.26\textwidth}
    \centering
    \includegraphics[width=\textwidth]{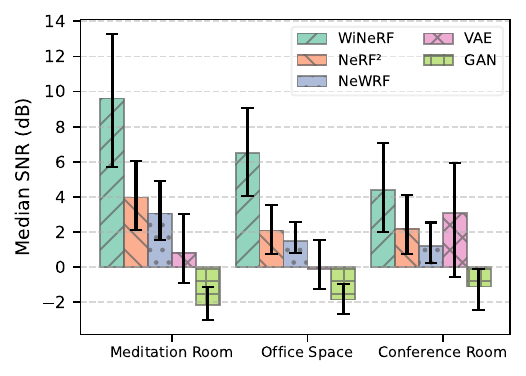}
    \caption{Prediction SNR per room \\ (Higher is better)}
    \label{fig:snr_bar}
  \end{subfigure}
  \begin{subfigure}[t]{0.26\textwidth}
    \centering
    \includegraphics[width=\textwidth]{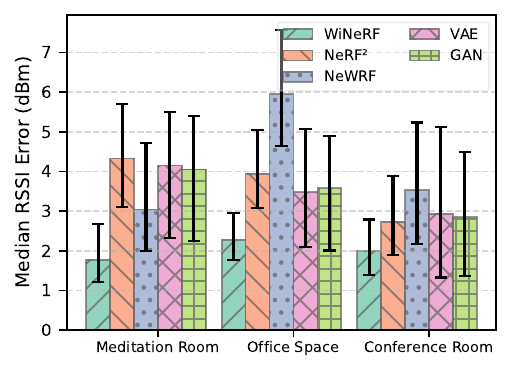}
    \caption{RSSI Estimation Error \\ (Lower is better)}
    \label{fig:rssi_bar}
  \end{subfigure}
  \begin{subfigure}[t]{0.26\textwidth}
    \centering
    \includegraphics[width=\textwidth]{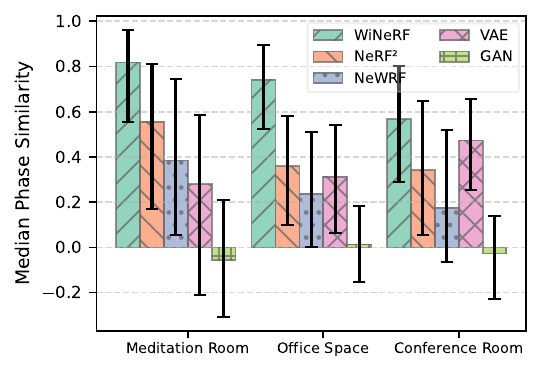}
    \caption{Phase Vector Similarity \\ (Higher is better)}
    \label{fig:phase_bar}
  \end{subfigure}
  \hspace*{-0.06\textwidth} 
  \caption{\sysname{} outperforms all baselines in channel estimation accuracy at unseen WiFi receiver locations, while accurately preserving both magnitude and phase characteristics.}
  
  \label{fig:comprehensive_evaluation}
\end{figure*}

\section{Evaluation}

\rev{We evaluate \sysname{} on the task of estimating the full 4$\times$50 complex CSI channel across the three indoor environments shown in Figure~\ref{fig:simulation_env_models}, selected to span different room sizes, geometries, and multipath conditions and thus assess site-specific reconstruction under diverse static layouts. Here, 4 denotes the number of receive antennas and 50 denotes the number of OFDM subcarriers per packet. This setup allows us to assess both spatial (across antennas) and frequency (across subcarriers) reconstruction fidelity, which is critical for practical MIMO applications such as beamforming, localization, and coverage mapping.}

\subsection{Baselines}

We compare our system's performance against the following state-of-the-art and classical baselines:
\begin{itemize}[topsep=0pt, itemsep=0pt, parsep=0pt, leftmargin=*]
    \item \textbf{NeRF$^2$}~\cite{zhao2023nerf2}, which models the spatial behavior of radio signals using a continuous volumetric representation parameterized by MLPs. The model can estimate complex CSI by accumulating contributions along the signal propagation path. We employ the publicly available implementation  and train it on our datasets. Since the framework is designed for scenarios with moving transmitters, we prepare compatible datasets by fixing device locations and interchanging the transmitter and receiver roles.
    \item \textbf{NeWRF}~\cite{lu2024deep}, which models wireless propagation as a volumetric field guided by direction-of-arrival (DoA) priors. As NeWRF requires DoA input, we estimate the top three arrival directions per receiver using the MUSIC algorithm on the ground truth CSI matrix and use them to train the model on our datasets.
    \item \textbf{VAE}~\cite{kingma2013auto}, a generative model. Variational Autoencoders have previously been used to generate wireless channel representations, albeit in different contexts~\cite{liu2021fire}.

    \item \textbf{DC-GAN}~\cite{goodfellow2014generative}, a deep convolutional GAN chosen as a baseline because it represents a generative modeling approach capable of learning complex, high-dimensional data distributions. This allows a direct comparison between a generative neural network and \sysname{}.

\end{itemize}

\subsection{Evaluation Metrics}
We evaluate \sysname{} using three complementary metrics that assess different aspects of channel estimation quality: overall channel fidelity, signal power estimation accuracy, and phase coherence preservation. These metrics collectively provide a comprehensive assessment of the model's ability to reconstruct physically meaningful and actionable channel representations.

\vspace{.3em}\noindent \textbf{Channel Prediction SNR:} This metric measures the overall fidelity of the complex channel estimate by computing the signal-to-noise ratio between predicted and ground-truth CSI matrices, and is employed by prior works to assess overall channel quality \cite{zhao2023nerf2, liu2021fire, luo2022learning}. The SNR is defined as:
\begin{equation}
\text{SNR}_{\text{pred}} = -10 \log_{10} \left( \frac{\|\mathbf{H}_{\text{pred}} - \mathbf{H}_{\text{true}}\|_2}{\|\mathbf{H}_{\text{true}}\|_2} \right) \quad \text{[dB]}
\end{equation}
where $\mathbf{H}_{\text{pred}}$ and $\mathbf{H}_{\text{true}}$ are the predicted and ground-truth channel matrices, respectively. Higher SNR values indicate better reconstruction quality, and, by construction, this metric captures both amplitude and phase accuracy.

\vspace{.3em}\noindent \textbf{RSSI Estimation Error:} This metric evaluates the accuracy of received signal strength by comparing the average power of predicted and measured channels. RSSI is computed from the CSI magnitude as:
\begin{equation}
\text{RSSI} = \frac{1}{N_{\text{ant}} N_{\text{sc}}} \sum_{i=1}^{N_{\text{ant}}} \sum_{j=1}^{N_{\text{sc}}} |H[i,j]|^2 \quad \text{[dBm]},
\end{equation}
where $N_{\text{ant}}$ and $N_{\text{sc}}$ denote the number of antennas and subcarriers. The RSSI error is the absolute difference between predicted and ground-truth values; lower errors indicate better power estimation, which is critical for coverage mapping and link budget analysis \cite{molisch2012wireless}.

\vspace{.3em}\noindent \textbf{Relative Phase Similarity:} This metric evaluates how well the model preserves spatial phase relationships across antenna elements, which is critical for applications such as beamforming and angle-of-arrival estimation. For each subcarrier, we normalize the phase by the first antenna to remove global phase offsets, then compute the cosine similarity between predicted and ground-truth phase vectors:
\begin{equation}
\text{Similarity} = \frac{\boldsymbol{\phi}_{\text{pred}} \cdot \boldsymbol{\phi}_{\text{true}}}{\|\boldsymbol{\phi}_{\text{pred}}\| \, \|\boldsymbol{\phi}_{\text{true}}\|},
\end{equation}
where $\boldsymbol{\phi}_{\text{pred}}$ and $\boldsymbol{\phi}_{\text{true}}$ are the phase vectors across antennas for a given subcarrier. Values range from $-1$ to $1$, with $1$ indicating perfect alignment. This metric quantifies the model's ability to capture the relative phase structure across antennas and subcarriers.

\subsection{Channel Estimation Accuracy}
\vspace{.3em} \noindent \textbf{SNR performance:} As shown in Figure~\ref{fig:snr_cdf}, \sysname{} achieves a median prediction SNR of 5.30~dB and 90th percentile of 11.29~dB, outperforming all baselines by at least 130\%. This is mainly because all baselines heavily rely on dense training data or priors. NeRF$^2$ requires a very dense sampling of \textasciitilde178~measurements/ft$^3$, while our experimental data has an average density of \textasciitilde4~measurements/ft$^3$. On the other hand, NeWRF heavily depends on precise angle-of-arrival (AoA) priors, which are not available in real-world measurements. Figure~\ref{fig:snr_bar} presents the prediction SNR across different environments, showing that \sysname{} consistently outperforms all baselines regardless of environment size or multipath complexity. In the smallest environment, the Meditation Room, \sysname{} achieves a median SNR of 9.61~dB, compared to 3.96~dB for NeRF$^2$ and 3.05~dB for NeWRF, yielding gains of 5.6–8.8~dB. While performance slightly decreases in larger environments due to increased spatial extent and richer multipath, \sysname{} still surpasses all baselines by a clear margin. Even in these more challenging settings, the resulting SNR remains sufficient to produce actionable and spatially consistent estimations across space and antenna elements, as we demonstrate in the following results.

\vspace{.3em} \noindent \textbf{RSSI Estimation Performance:} While the previous analysis focused on SNR, which captures both the magnitude and phase similarity of the predicted complex channel, we now examine the role of CSI magnitude in RSSI estimation. RSSI serves as a key indicator for coverage mapping and MCS index selection in WiFi systems. Figure~\ref{fig:rssi_bar} presents the per-room RSSI error, showing that \sysname{} achieves a median RSSI error between 1.8 and 2.3~dB, consistently outperforming all baselines by 35–50\% and delivering more accurate power estimations across diverse environments.

\vspace{.3em} \noindent \textbf{Channel Phase Similarity Performance: }
Preserving relative phase across antennas and subcarriers is essential for maintaining multipath information. To evaluate this capability, we compute the similarity between the estimated and ground-truth relative phase vectors. Figure~\ref{fig:phase_bar} shows the per-room phase similarity, where \sysname{} achieves a median phase similarity between 0.5 and 0.8, outperforming all baselines by 40–300\%. \sysname{}’s performance in relative phase estimation slightly drops in the largest room due to numerous NLoS cases. We next show how these similarity factors are reflected in AoA estimation. %

\begin{figure}[t]
  \hspace*{-0.06\textwidth} 
  \centering
  \begin{subfigure}[t]{0.54\columnwidth}
    \centering
    \includegraphics[width=\textwidth]
    {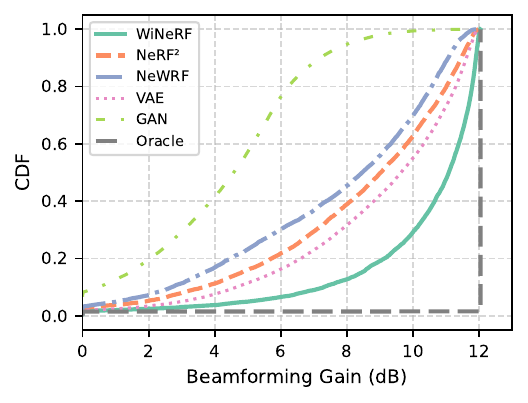}
    \caption{Beamforming Gain \\ (Higher is better)}
    \label{fig:beamforming_cdf}
  \end{subfigure}
  \begin{subfigure}[t]{0.54\columnwidth}
    \centering
    \includegraphics[width=\textwidth]{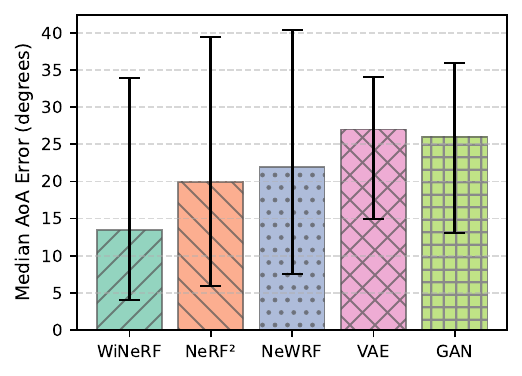}
    \caption{\gls{los} Angle-of-Arrival Estimation Error (Lower is better)}
    \label{fig:aoa_bar}
  \end{subfigure}
  \hspace*{-0.06\textwidth} 
  \caption{\sysname{} preserves phase coherence for practical MIMO tasks: (a) Beamforming and (b) \gls{aoa} estimation, using predicted CSI with classical signal processing methods.}
  \label{fig:aoa_beamforming}
\end{figure}

\subsection{Beamforming Gain Performance}
\label{subsec:beamforming-egc}

We validate that WiNeRF’s channel estimates are not only numerically accurate but also physically actionable, preserving the phase coherence necessary for beamforming. To test this, we perform equal-gain combining (EGC), a stringent benchmark that relies solely on CSI phase information using the estimated CSI for test data. In EGC, the combining weights are defined as \(w_{m,s} = e^{-j \angle h_{m,s}}\), meaning any phase error directly causes destructive interference and loss of array gain. We use our test data and the estimated CSI across 4 antennas in all three environments and compute the beamforming gain as
\begin{equation}
    G_s = 10 \log_{10}\!\left( \frac{P_{\text{EGC},s}}{P_{\text{avg},s}} \right),
\end{equation}
aggregated over 50 subcarriers per packet. Importantly, the combining weights are derived from the predicted CSI but applied to the measured channels, isolating phase accuracy from other confounding factors.
Figure~\ref{fig:beamforming_cdf} presents the CDFs of beamforming gains. WiNeRF achieves a median gain of 11.4 dB, closely tracking the oracle’s 12 dB (only a 0.6 dB or 5\% gap), while outperforming all baselines by large margins, up to 145\% higher than NeRF$^2$ and even larger improvements over NeWRF, VAE, and GAN. Unlike these models, WiNeRF’s gain distribution is consistently high, with fewer than 10\% of packets falling below 8 dB, compared to about 40\% for baselines.

In practical terms, this 3 dB improvement in beamforming translates to a 19\% increase in theoretical link capacity for a 20 MHz Wi-Fi channel at 15 dB baseline SNR, equivalent to a doubling of range or a fourfold reduction in transmit power. These results demonstrate that \sysname{}’s CSI estimates retain the correct relative phase structure across antennas and subcarriers, making them immediately usable for beamforming and other downstream PHY-layer operations without requiring training of a new downstream task.

\subsection{\gls{los} Angle-of-Arrival Estimation Accuracy}
\label{subsec:aoa-estimation}
\sysname{} is designed to preserve the underlying physics of wireless propagation, particularly multipath structure such as the \gls{aoa}, which is captured in the relative phase across antennas. The \gls{aoa} of the direct path between the WiFi transmitter and receiver is especially important for applications such as localization, tracking, and robotic path planning. To validate that \sysname{}’s channel estimates retain this physical information, we apply the classical MUSIC algorithm \cite{kotaru2015spotfi, schmidt1986multiple} to the predicted CSI and evaluate its ability to recover the \gls{los} \gls{aoa} at unseen receiver locations \rev{within the same environment.}.

\begin{figure}[t]
    \centering
    \includegraphics[width=\linewidth,trim={0.1cm 0.1cm 0.1cm 0.1cm},clip]{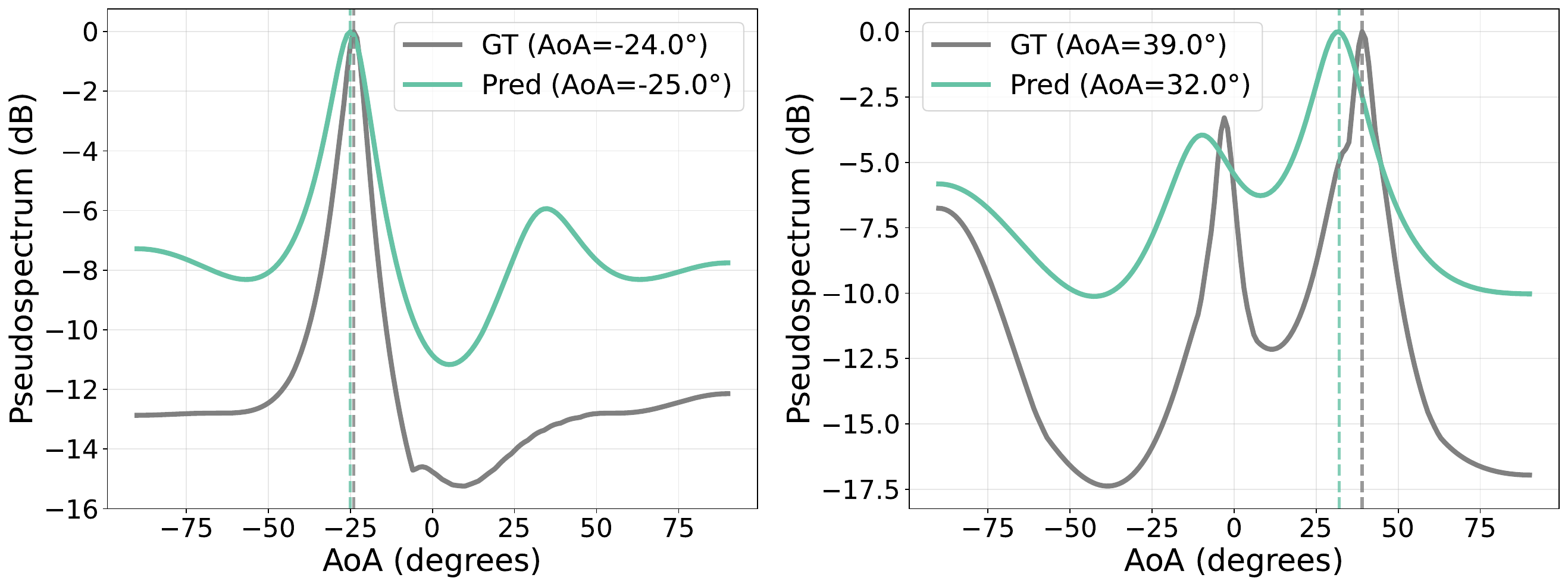} 
    \caption{Multipath resolution results using estimated CSI; \sysname{} preserves accurate multipath AoA characterization}
    \label{fig:Music_Spectrum}
\end{figure}

Figure~\ref{fig:aoa_bar} shows the \gls{los} \gls{aoa} estimation error across methods. \sysname{} achieves a median error of 13.0°, consistently outperforming all baselines. In practical terms, this corresponds to an angular bias of roughly 11\% within a 120° sector, translating to a localization error of about 23 cm at a 1m distance that is sufficient for coarse, room-scale tracking without additional model training. As illustrated in Figure~\ref{fig:Music_Spectrum}, \sysname{}’s predicted CSI produces sharp, well-aligned LoS and multipath peaks. This ability to maintain spatial structure makes \sysname{}’s CSI actionable for beam tracking and interference localization. It is worth noting that unlike prior work such as NeRF$^2$, which trains a separate downstream network to infer spatial spectra from synthetic wireless data generated by a base NeRF model, \sysname{} directly enables classical signal processing methods (e.g., MUSIC) to recover multipath characteristics from its predicted CSI. This distinction demonstrates that \sysname{} truly captures the underlying physical multipath structure rather than merely memorizing patterns inside images.

\subsection{Radio Coverage Prediction via RSSI Mapping}
\label{subsec:rssi-coverage}
\begin{figure}[t]
    \centering
    \includegraphics[width=0.95\linewidth]{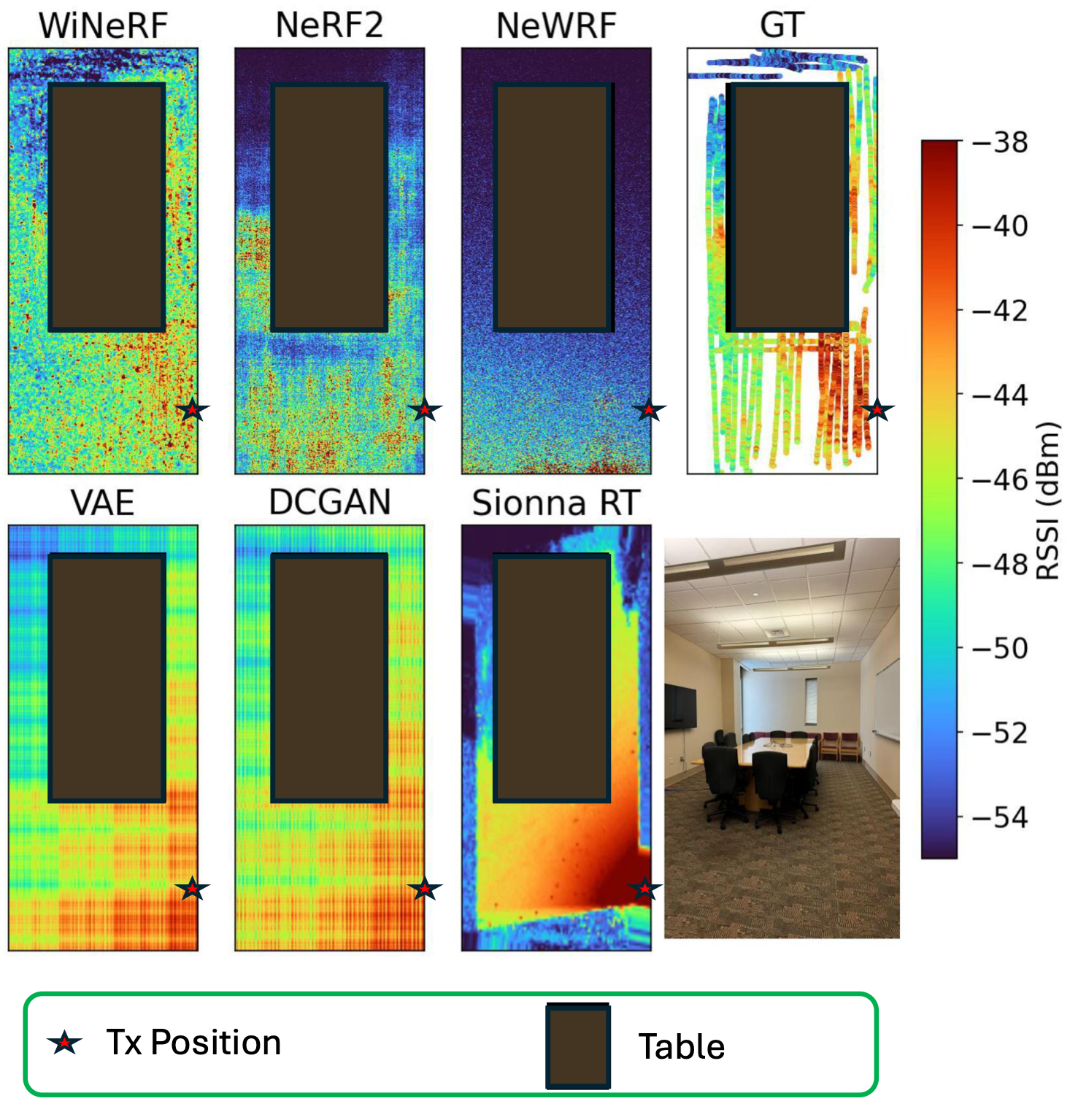} 
    \caption{RSSI spatial coverage prediction in Conference Room. WiNeRF accurately reproduces room-scale coverage patterns with fine spatial detail matching ground truth, capturing both the LoS hotspot and NLoS attenuation at edges.} %
    \label{fig:RSSI_Heatmap}
\end{figure}

Next, we demonstrate \sysname{}’s ability to preserve spatial CSI patterns across an entire room. This is an essential capability for applications such as coverage mapping, access point (AP) placement, link budgeting, and trajectory planning. Using pretrained CSI estimation models from \sysname{} and the baselines, we query each model over a dense 2D spatial grid in the conference room and convert the predicted CSI into RSSI values. Figure~\ref{fig:RSSI_Heatmap} compares the inferred coverage maps with sparse ground-truth measurements, showing that \sysname{} accurately captures the expected spatial variations, including RSSI drops in NLoS regions and constructive or destructive multipath interference patterns. Quantitatively, \sysname{} achieves lower median RSSI error than the strongest learned baseline and also reduces worst-case error, indicating that its spatial predictions are not only visually plausible but also more reliable for deployment-facing tasks such as dead-zone identification and path planning. This results in actionable, physically consistent predictions suitable for proactive, context-aware decision-making.

For comparison, Figure~\ref{fig:RSSI_Heatmap} also includes a ray-tracing--based coverage map generated using SionnaRT and a detailed 3D CAD model of the same environment. While ray tracing can yield precise results when surface materials and geometry are modeled accurately, its performance degraded around the conference table due to incomplete geometry and missing details beneath the table captured during scanning. This comparison further highlights that \sysname{} can recover room-scale coverage structure directly from sparse measurements even when complete geometric fidelity is unavailable.

\begin{figure*}[t]
  \hspace*{-0.06\textwidth} 
  \centering
  \begin{subfigure}[t]{0.26\textwidth}
    \centering
    \includegraphics[width=\textwidth]{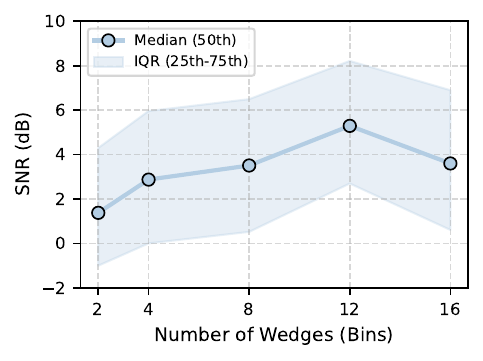}
    \caption{Impact of Cone Sizes}
    \label{fig:sensitivity_num_cones}
  \end{subfigure}
  \begin{subfigure}[t]{0.26\textwidth}
    \centering
    \includegraphics[width=\textwidth]
    {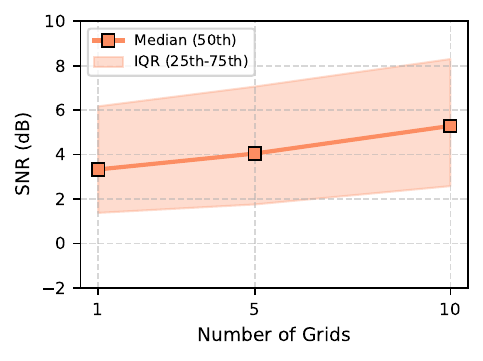}
    \caption{Impact of Hash Grid Levels}

    \label{fig:sensitivity_num_grids}
  \end{subfigure}
  \begin{subfigure}[t]{0.26\textwidth}
    \centering
    \includegraphics[width=\textwidth]{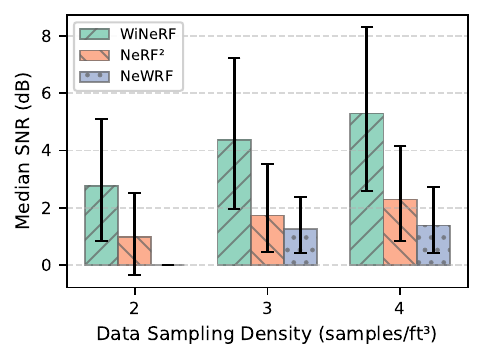}
    \caption{Impact of Training Data Density}
    \label{fig:sensitivity_data_density}
  \end{subfigure}
  \begin{subfigure}[t]{0.26\textwidth}
    \centering
    \includegraphics[width=\textwidth]
    {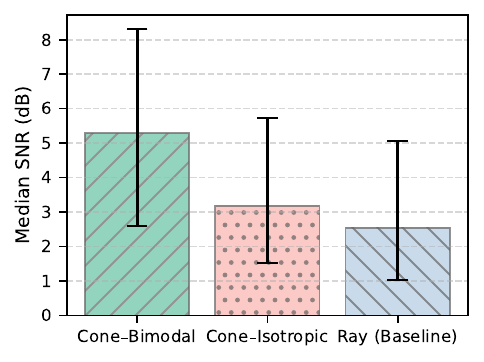}
    \caption{Ablation Study of Cone Type}
    \label{fig:ablation_cone_type}
  \end{subfigure}
    \hspace*{-0.06\textwidth} 
  \caption{\sysname{} Sensitivity Analysis}
  \label{fig:independent_metrics}
\end{figure*}

\section{Sensitivity Analysis}
Next, we evaluate \sysname{}’s sensitivity to key design factors, including cone geometry, hash-grid parameters, and training data density. We further conduct an ablation study to quantify the contribution of the proposed cone-based ray structure.

\vspace{.3em} \noindent \textbf{Impact of Cone Sizes:}
We evaluate how the angular discretization in our wave model influences channel estimation accuracy by varying the number of conical rays around each receiver. This parameter defines the trade-off between spatial resolution and computational efficiency in the cone-based representation. We test configurations with 2–16 cones, corresponding to azimuthal bin widths from 180° to 22.5°, while keeping all other model parameters fixed. As shown in Fig.~\ref{fig:sensitivity_num_cones}, performance improves with finer angular sampling up to 12 cones, which yields the highest median prediction SNR of 5.3~dB. This aligns with the theoretical angular resolution limit imposed by our 4-antenna array hardware.

\vspace{.3em} \noindent \textbf{Impact of Hash Grid Levels:} We analyze how the number of multi-resolution grid levels influences channel estimation accuracy (Fig.~\ref{fig:sensitivity_num_grids}). The grids span spatial resolutions from coarse (meter-scale) to fine (centimeter-scale), and we vary the number of levels used to represent this range. Results show a clear upward trend in performance as grid depth increases, confirming that distributing the representation across multiple exponentially spaced levels enables the model to capture both large-scale spatial trends and fine-grained multipath variations. This finding underscores the value of hierarchical feature encoding for accurately modeling complex wireless propagation environments.

\vspace{.3em} \noindent \textbf{Impact of Training Data Density: }
We evaluate how training data density influences channel estimation accuracy across all environments (Fig.~\ref{fig:sensitivity_data_density}). As expected, higher sampling density improves performance for all models. However, \sysname{} consistently achieves higher SNR and degrades more gracefully under sparse data conditions, whereas baseline methods show sharp performance drops at lower densities. \rev{This robustness highlights \sysname{}’s efficiency in learning spatial channel structure from limited training data and predicting at held-out receiver locations within the same environment.}.

\vspace{.3em} \noindent \textbf{Ablation Study:}
We compare three spatial propagation modeling strategies to assess their impact on channel estimation performance (Fig.~\ref{fig:ablation_cone_type}): (1) our proposed anisotropic wedge cone, \rev{which uses wedge-shaped sampling regions to reflect the directional angular observability constraints of the antenna array;} (2) an isotropic cone, which radiates uniformly in all directions; and (3) a simple ray baseline, which uses idealized line rays without spatial spreading, similar to NeRF$^2$~\cite{zhao2023nerf2} and NeWRF~\cite{lu2024deep}. All configurations employ 12 angular bins (30° azimuth resolution) and are evaluated across three environments, with the ray baseline using 12 rays centered in each bin. As shown in Fig.~\ref{fig:ablation_cone_type}, the anisotropic wedge cone achieves the highest median SNR of 5.29~dB, demonstrating that incorporating \rev{angular anisotropy consistent with the array’s measurement constraints} improves ULA-based channel estimation over both omnidirectional and idealized ray models.

\section{Discussion and Future Works}

\sysname{} effectively learns both phase coherence and multipath characteristics, enabling physically meaningful and actionable channel representations. \Raf{By embedding hardware-informed inductive biases directly into the neural field, the framework successfully reconstructs complex CSI even under the tight sensing and communication constraints typical of embedded platforms.} \rev{However, the current paper evaluates \sysname{} only in static environments and therefore does not establish performance under moving scatterers and changing layouts.} While the current framework focuses on linear antenna arrays, it is readily extendable to other antenna geometries.

\Raf{We intentionally designed \sysname{} as a site-specific model to support high-fidelity Digital Twins and spatial planning tools, where capturing the unique multipath "fingerprint" of an environment is the primary objective.} This specialization allows the model to achieve high prediction accuracy in complex, static scenes from sparse measurements, \rev{but it also means that \sysname{} is currently trained separately for each environment and does not include a cross-environment transfer or rapid adaptation mechanism.} \rev{At present, \sysname{} is evaluated as a server-side system for digital twinning applications, and embedded deployment remains an important direction for future work.}

\rev{Several extensions remain open. First, dynamic environments with moving objects are not validated by the current experiments. A possible future direction is to incorporate object-aware neural representations together with external physical priors such as camera or radar feeds to help disentangle static background structure from dynamic components. Second, transfer learning or related adaptation strategies could reduce retraining cost and improve portability across environments. }%

\section{Conclusion}

We presented \sysname{}, a neural wireless channel modeling framework that learns a spatially continuous channel representation constrained by what is observable from commodity wireless hardware, without relying on idealized thin-ray propagation or external priors such as known geometry or explicit \gls{aoa} information. By incorporating typical constraints of embedded platforms such as antenna geometry and finite angular resolution into its modeling formulation, \sysname{} focuses on learning the subset of channel variations that can be reliably inferred from sparse and noisy measurements. This constrained representation enables stable and reusable channel estimates that can be directly queried for downstream tasks.
Importantly, the learned CSI can be used with standard signal-processing pipelines—such as beamforming, \gls{los} \gls{aoa} estimation, and signal coverage mapping—without task-specific retraining or changes to existing hardware or protocols. Experimental evaluation across three indoor environments demonstrates a median prediction SNR of 5.30 dB (3.0–6.7 dB improvement over prior methods), near-oracle beamforming performance with only 0.6 dB loss, and a median \gls{los} \gls{aoa} error of 13°. Together, these results show that \sysname{} provides a practical and task-agnostic channel representation that supports proactive, wireless-aware decision making in embedded \gls{isac} systems.

\begin{acks}
We thank Mr.~Jakob Link and Dr.~Matthias Hollick from the Secure Mobile Networking Lab at TU Darmstadt for their help with updating the router firmware files used in our data collection pipeline. We also thank Navid Tajkhorshid, an undergraduate student at the University of Illinois Urbana-Champaign, for his help with setting up the data collection system.
This work was supported in part by the National Science Foundation (NSF) under Grants~\#2414227, and by Intel. %
\end{acks}

\bibliographystyle{ACM-Reference-Format}
\bibliography{Sections/reference}

\end{document}